\documentclass[final            
]{aipprocdlw}

\layoutstyle{8x11single}

\begin{document}
\font\fiverm=cmr5\font\sevenrm=cmr7\font\sevenit=cmmi7
\def\sbf{\footnotesize\bf}
\def\ns#1{_{\hbox{\sevenrm #1}}}
\def\PRL#1{Phys.\ Rev.\ Lett.\ {\bf#1}} \def\PR#1{Phys.\ Rev.\ {\bf#1}}
\def\ApJ#1{Astrophys.\ J.\ {\bf#1}}
\def\AsJ#1{Astron.\ J.\ {\bf#1}}
\def\MNRAS#1{Mon.\ Not.\ R.\ Astr.\ Soc.\ {\bf#1}}
\def\CQG#1{Class.\ Quantum Grav.\ {\bf#1}}
\def\GRG#1{Gen.\ Relativ.\ Grav.\ {\bf#1}}
\def\beq{\begin{equation}} \def\eeq{\end{equation}}
\def\bea{\begin{eqnarray}} \def\eea{\end{eqnarray}}
\def\Z#1{_{\lower2pt\hbox{$\scriptstyle#1$}}} \def\w#1{\,\hbox{#1}}
\def\X#1{_{\lower2pt\hbox{$\scriptscriptstyle#1$}}}
\font\sevenrm=cmr7 \def\ns#1{_{\hbox{\sevenrm #1}}} \def\dOM{\dd\Omega^2}
\def\Ns#1{\Z{\hbox{\sevenrm #1}}} \def\ave#1{\langle{#1}\rangle}
\def\lsim{\mathop{\hbox{${\lower3.8pt\hbox{$<$}}\atop{\raise0.2pt\hbox{$\sim$}}
$}}} \def\kmsMpc{\w{km}\;\w{sec}^{-1}\w{Mpc}^{-1}} \def\bn{\bar n}
\def\dd{{\rm d}} \def\ds{\dd s} \def\etal{{\em et al}.}
\def\al{\alpha}\def\be{\beta}\def\ga{\gamma}\def\de{\delta}\def\ep{\epsilon}
\def\et{\eta}\def\th{\theta}\def\ph{\phi}\def\rh{\rho}\def\si{\sigma}
\def\gsim{\mathop{\hbox{${\lower3.8pt\hbox{$>$}}\atop{\raise0.2pt\hbox{$
\sim$}}$}}} \def\ta{\tau} \def\ac{a}
\def\frn#1#2{{\textstyle{#1\over#2}}} \def\Deriv#1#2#3{{#1#3\over#1#2}}
\def\Der#1#2{{#1\hphantom{#2}\over#1#2}} \def\pt{\partial} \def\ab{{\bar a}}
\def\tw{\ta}\def\gb{\bar\ga} \def\BB{{\cal B}} \def\CC{{\cal C}}
\def\av{{a\ns{v}\hskip-2pt}} \def\aw{{a\ns{w}\hskip-2.4pt}}\def\Vav{{\cal V}}
\def\DD{{\cal D}}\def\gd{{{}^3\!g}}\def\half{\frn12}\def\Rav{\ave{\cal R}}
\def\QQ{{\cal Q}}\def\dsp{\displaystyle} \def\rw{r\ns w}
\def\mean#1{{\vphantom{\tilde#1}\bar#1}} \def\bx{{\mathbf x}}
\def\bH{\mean H}\def\Hb{\bH\Z{\!0}}\def\bq{\mean q}
\def\gb{\mean\ga}\def\gc{\gb\Z0} \def\OMMn{\OM\Z{M0}}
\def\rhb{\mean\rh}\def\OM{\mean\Omega}\def\etb{\mean\eta}
\def\fw{{f\ns w}}\def\fv{{f\ns v}} \def\goesas{\mathop{\sim}\limits}
\def\fvn{f\ns{v0}} \def\fvf{\left(1-\fv\right)} \def\Hh{H}
\def\OMM{\OM\Z M}\def\OmMn{\Omega\Z{M0}} \def\ts{t}\def\tb{\ts'}  \def\Hm{H\Z0}
\def\Fi{\hbox{\footnotesize\it fi}}\def\etw{\eta\ns w} \def\etv{\eta\ns v}
\def\fvi{{f\ns{vi}}} \def\fwi{{f\ns{wi}}}
\def\Hx{H_{\lower1pt\hbox{$\scriptstyle0$}}}
\def\LCDM{$\Lambda$CDM} \def\OmLn{\Omega\Z{\Lambda0}}\def\Omkn{\Omega\Z{k0}}
\def\Dtc{\mathop{\hbox{$\Der\dd\tw$}}}
\def\tbA{\left(2t+3b\right)} \def\tbB{\left(2t^2+3bt+2b^2\right)}
\def\Dfb{D\Ns{TS}} \def\Dlcdm{D\Ns{$\scriptstyle\Lambda$CDM}}
\def\dL{d\Z L} \def\dA{d\Z A} \def\name{timescape}
\def\OmBn{\Omega\Z{B0}} \def\OmCn{\Omega\Z{C0}} 
\title{Gravitational energy as dark energy: Average observational quantities$^*$}
\author{David L. Wiltshire}{
address={Department of Physics \& Astronomy, University of Canterbury, Private Bag 4800, Christchurch 8140,\\ New Zealand; and}
,altaddress={International Center for Relativistic Astrophysics Network (ICRANet), P.le della Repubblica 10, Pescara 65121, Italy}
}
\thispagestyle{addnote}
\begin{abstract}
In the timescape scenario cosmic acceleration is understand as an apparent
effect, due to gravitational energy gradients that grow when spatial
curvature gradients become significant with the nonlinear growth of cosmic
structure. This affects the calibratation of local geometry to the
solutions of the volume--average evolution equations corrected by
backreaction. In this paper I discuss recent work on defining observational
tests for average geometric quantities which can distinguish the timescape
model from a cosmological constant or other models of dark energy.
\end{abstract}

\maketitle
\section{Introduction}

I will discuss some recent results on observational tests \cite{obs}
of a model cosmology, which represents a new approach to understanding
the phenomenology of dark energy as a consequence of the effect of
the growth of inhomogeneous structures. The basic idea, outlined in
a nontechnical manner in ref.\ \cite{dark07}, is that as inhomogeneities
grow one must consider not only their backreaction on average cosmic
evolution -- as discussed by other contributors to this volume -- but
also the variance in the geometry as it affects the
calibration of clocks and rods of ideal observers. Dark energy is
then effectively realised as a misidentification of gravitational
energy gradients.

Although the standard Lambda Cold Dark Matter (\LCDM) model provides a good
fit to many tests, there are tensions between some tests, and also a number
of puzzles and anomalies. Furthermore, at the present epoch the observed
universe is only statistically homogeneous once one samples on scales
of 150--300 Mpc. Below such scales it displays a web--like structure,
dominated in volume by voids. Some 40\%--50\% of the volume of the present
epoch universe is in voids with $\de\rho/\rho\goesas-1$ on scales of
30$h^{-1}$ Mpc \cite{HV}, where $h$ is the dimensionless parameter related to
the Hubble constant by $H\Z0=100h\kmsMpc$. Once one also accounts for
numerous minivoids, and perhaps also a few larger voids, then it appears that
the present epoch universe is void-dominated. Clusters of galaxies are spread
in sheets that surround these voids, and in thin filaments that thread them.

One particular consequence of a matter distribution that is only
statistically homogeneous, rather than exactly homogeneous, is that when
the Einstein equations are averaged they do not evolve as a smooth
Friedmann--Lema\^{\i}tre--Robertson--Walker (FLRW) geometry. Instead
the Friedmann equations are supplemented by additional backreaction
terms \cite{buch00}. Whether or not one can fully explain the expansion
history of the universe as a consequence of the growth of inhomogeneities
and backreaction, without a fluid--like dark energy, is the subject of
ongoing debate \cite{buch08}.

Elsewhere in this volume, Peebles \cite{Peebles} provides some
of the arguments that have been presented against backreaction. His
line of reasoning is that of a plausibility argument: if we {\em assume}
a FLRW geometry with small perturbations, and estimate the magnitude of
the perturbations from the typical rotational and peculiar velocities of
galaxies, then the corrections of inhomogeneities are consistently small. This
would be a powerful argument, were it not for the fact that at the present
epoch galaxies are not homogeneously distributed. The Hubble Deep Field
reveals that galaxies were close to being homogeneous distributed at early
epochs, but following the growth voids at redshifts $z\lsim1$
that is no longer the case today. Therefore galaxies cannot be consistently
treated as randomly distributed gas particles on the 30$h^{-1}$ Mpc scales
\cite{HV} that dominate present cosmic structure below the scale of
statistical homogeneity.

Over the past few years I have developed a new physical interpretation of
cosmological solutions within the Buchert averaging scheme
\cite{clocks,sol,equiv}. I start
by noting that in the presence of strong spatial curvature
gradients, not only should the average evolution equations be replaced
by equations with terms involving backreaction, but the physical
interpretation of average quantities must also account for the differences
between the local geometry and the average geometry.
In other words, geometric variance can be just as important as
geometric averaging when it comes to the physical interpretation of the
expansion history of the universe. 

I proceed from the fact that structure formation provides a natural division
of scales in the observed universe. As observers in galaxies, we and the
objects we observe in other galaxies are necessarily in bound structures,
which formed from density perturbations that were greater than critical
density. If we consider the evidence of the large scale structure
surveys on the other hand, then the average location by volume in the
present epoch universe is in a void, which is negatively curved.
We can expect systematic
differences in spatial curvature between the average mass environment, in
bound structures, and the volume-average environment, in voids.

Spatial curvature gradients will in general give rise to gravitational
energy gradients. Physically this can be understood in terms of a
relative deceleration of expanding regions of different densities.
Those in the denser region decelerate more and age less. Since we
are dealing with weak fields the relative deceleration of the background
is small. Nonetheless even if the relative deceleration is typically of
order $10^{-10}$ms$^{-2}$, cumulatively over the age of the universe
it leads to significant clock rate variances \cite{equiv}. I proceed
from an ansatz that the variance in gravitational energy is correlated
with the average spatial curvature in such a way as to implicitly
solve the Sandage--de Vaucouleurs paradox that a statistically quiet,
broadly isotropic, Hubble flow is observed deep below the scale of
statistical homogeneity. In particular, galaxy peculiar velocities
have a small magnitude with respect to a local regional volume expansion.
Expanding regions of different densities are patched together so that the
regionally measured expansion remains uniform. Such regional expansion
refers to the variation of the regional
proper length, $\ell_r=\Vav^{1/3}$, with respect to proper time of
isotropic observers (those who see an isotropic mean CMB).
Although voids open up faster, so that their proper volume increases more
quickly, on account of gravitational energy gradients the local clocks will
also tick faster in a compensating manner.

Details of the fitting of local observables to average quantities for
solutions to the Buchert formalism are described in detail in refs.\
\cite{clocks,sol}. Negatively curved voids, and
spatially flat expanding wall regions within which galaxy clusters are
located, are combined in a Buchert average
\beq\fv(t)+\fw(t)=1,\eeq
where $\fw(t)=\fwi\aw^3/\ab^3$ is the {\em wall volume fraction} and
$\fv(t)=\fvi\av^3/\ab^3$ is the {\em void volume fraction},
$\Vav=\Vav\ns i\ab^3$ being the present horizon volume, and $\fwi$, $\fvi$ and
$\Vav\ns i$ initial values at last scattering. The time parameter, $t$,
is the volume--average time parameter of the Buchert formalism, but does not
coincide with that of local measurements in galaxies. In trying to fit a
FLRW solution to the universe we attempt to
match our local spatially flat wall geometry
\beq\ds^2\Z{\Fi}=-\dd\tw^2+\aw^2(\tw)\left[\dd\etw^2+
\etw^2\dOM\right]\,.
\label{wgeom}\eeq
to the whole universe, when in reality the rods and clocks of ideal isotropic
observers vary with gradients in spatial curvature and gravitational energy.
By conformally matching radial null geodesics with those of the Buchert
average solutions, the geometry (\ref{wgeom}) may be extended to cosmological
scales as the dressed geometry
\beq
\ds^2=-\dd\tw^2+\ac^2(\tw)\left[\dd\etb^2+\rw^2(\etb,\tw)\,\dOM\right]
\label{dgeom}\eeq
where $a=\gb^{-1}\ab$, $\gb=\Deriv\dd\tw\ts$ is the relative lapse
function between wall clocks and volume--average ones, $\dd\etb=\dd t/\ab=
\dd\tw/ \ac$, and $\rw=\gb\fvf^{1/3}\fwi^{-1/3}\etw(\etb,\tw)$, where
$\etw$ is given by integrating $\dd\etw=\fwi^{1/3}\dd\etb/[\gb\fvf^{1/3}]$
along null geodesics.

In addition to the bare cosmological parameters which describe the Buchert
equations, one obtains dressed parameters relative to the geometry
(\ref{dgeom}). For example, the dressed matter density parameter is
$\Omega\Z M=\gb^3\OMM$, where $\OMM=8\pi G\rhb\Z{M0}\ab\Z0^3/(3\bH^2\ab^3)$
is the bare matter density parameter. The dressed parameters take numerical
values close to the ones inferred in standard FLRW models.

\section{Apparent acceleration and Hubble flow variance}

The gradient in gravitational energy and cumulative differences of clock
rates between wall observers and volume average ones has important
physical consequences. Using the exact solution
obtained in ref.\ \cite{sol}, one finds that a volume average observer
would infer an effective deceleration parameter $\bq=-\ddot\ab/(\bH^2\ab)=
2\fvf^2/(2+\fv)^2$, which is always positive since there is no global
acceleration. However, a wall observer infers a dressed deceleration
parameter
\beq
q=-{1\over H^2 a}{\dd^2 a\over\dd\tw^2}=
{-\fvf(8\fv^3+39\fv^2-12\fv-8)\over\left(4+\fv+4\fv^2\right)^2}\,,
\label{qtrack}\eeq
where the dressed Hubble parameter is given by
\beq\Hh=\ac^{-1}\Dtc\ac=\gb\bH-\dot\gb\,=\gb\bH-\gb^{-1}\Dtc\gb\,.
\label{42}\eeq
At early times when $\fv\to0$ the dressed
and bare deceleration parameter both take the Einstein--de Sitter value
$q\simeq\bq\simeq\half$. However, unlike the bare parameter which
monotonically decreases to zero, the dressed parameter becomes negative
when $\fv\simeq0.59$ and $\bq\to0^-$ at late times. For the best-fit
parameters \cite{LNW} the apparent acceleration begins
at a redshift $z\simeq0.9$.

Cosmic acceleration is thus revealed as an apparent effect which arises
due to the cumulative clock rate variance of wall observers relative to
volume--average observers. It becomes significant only when the voids
begin to dominate the universe by volume. Since the epoch of onset of
apparent acceleration is directly related to the void fraction, $\fv$, this
solves one cosmic coincidence problem.

In addition to apparent cosmic acceleration, a second important apparent
effect will arise if one considers scales below that of statistical
homogeneity. By any one set of clocks it will appear that voids expand
faster than wall regions. Thus a wall observer will see galaxies on the
far side of a dominant void of diameter $30h^{-1}$ Mpc recede at a
rate greater than the dressed global average $\Hm$, while galaxies within
an ideal wall will recede at a rate less than $\Hm$. Since the uniform
bare rate $\bH$ would also be the local value within an ideal wall, eq.\
(\ref{42}) gives a measure of the variance in the apparent Hubble flow.
The best-fit parameters \cite{LNW} give a dressed Hubble
constant $\Hm=61.7^{+1.2}_{-1.1}\kmsMpc$, and a bare Hubble constant
$\Hb=48.2^{+2.0}_{-2.4}\kmsMpc$. The present epoch variance is 17--22\%.

Since voids dominate the universe by volume at the present epoch, any
observer in a galaxy in a typical wall region will measure locally higher
values of the Hubble constant, with peak values of order $72\kmsMpc$ at the
$30h^{-1}$ Mpc scale of the dominant voids. Over larger distances, as the
line of sight intersects more walls as well as voids, a radial spherically
symmetric average will give an average Hubble constant whose value decreases
from the maximum at the $30h^{-1}$ Mpc scale to the dressed global average
value, as the scale of homogeneity is approached at roughly the baryon
acoustic oscillation (BAO) scale of
$110h^{-1}$Mpc. This predicted effect could account for the Hubble bubble
\cite{JRK} and more detailed studies of the scale
dependence of the local Hubble flow \cite{LS}.

In fact, the variance of the local Hubble flow below the scale of homogeneity
should correlate strongly to observed structures in a manner which has no
equivalent prediction in FLRW models.

\section{Future observational tests}

There are two types of potential cosmological tests that can be developed;
those relating to scales below that of statistical homogeneity as discussed
above, and those
that relate to averages on our past light cone on scales much greater than
the scale of statistical homogeneity. The second class of tests includes
equivalents to all the standard cosmological tests of the standard FLRW model
with Newtonian perturbations. This second class of tests can be
further divided into tests which just deal with the bulk cosmological
averages (luminosity and angular diameter distances etc), and those that
deal with the variance from the growth of structures (late epoch integrated
Sachs--Wolfe effect, cosmic shear, redshift space distortions etc). Here
I will concentrate solely on the simplest tests which are directly related
to luminosity and angular diameter distance measures.

In the \name\ cosmology we have an effective dressed luminosity distance
\beq\dL=a\Z0(1+z)\rw,\eeq where $a\Z0=\gc^{-1}\ab\Z0$, and
\beq\rw=\gb\fvf^{1/3}
\int_\ts^{\ts\X0}{\dd\tb\over\gb(\tb)(1-\fv(\tb))^{1/3}\ab(\tb)}\,.
\label{eq:dL}\eeq
We can also define an {\em effective angular diameter distance}, $\dA$, and an
{\em effective comoving distance}, $D$, to a redshift $z$ in the
standard fashion
\beq\dA={D\over1+z}={\dL\over(1+z)^2}\,.\label{dist}\eeq

\begin{figure}
\centerline{{\sbf(a)}\hskip-15pt
\includegraphics[width=2.25in]{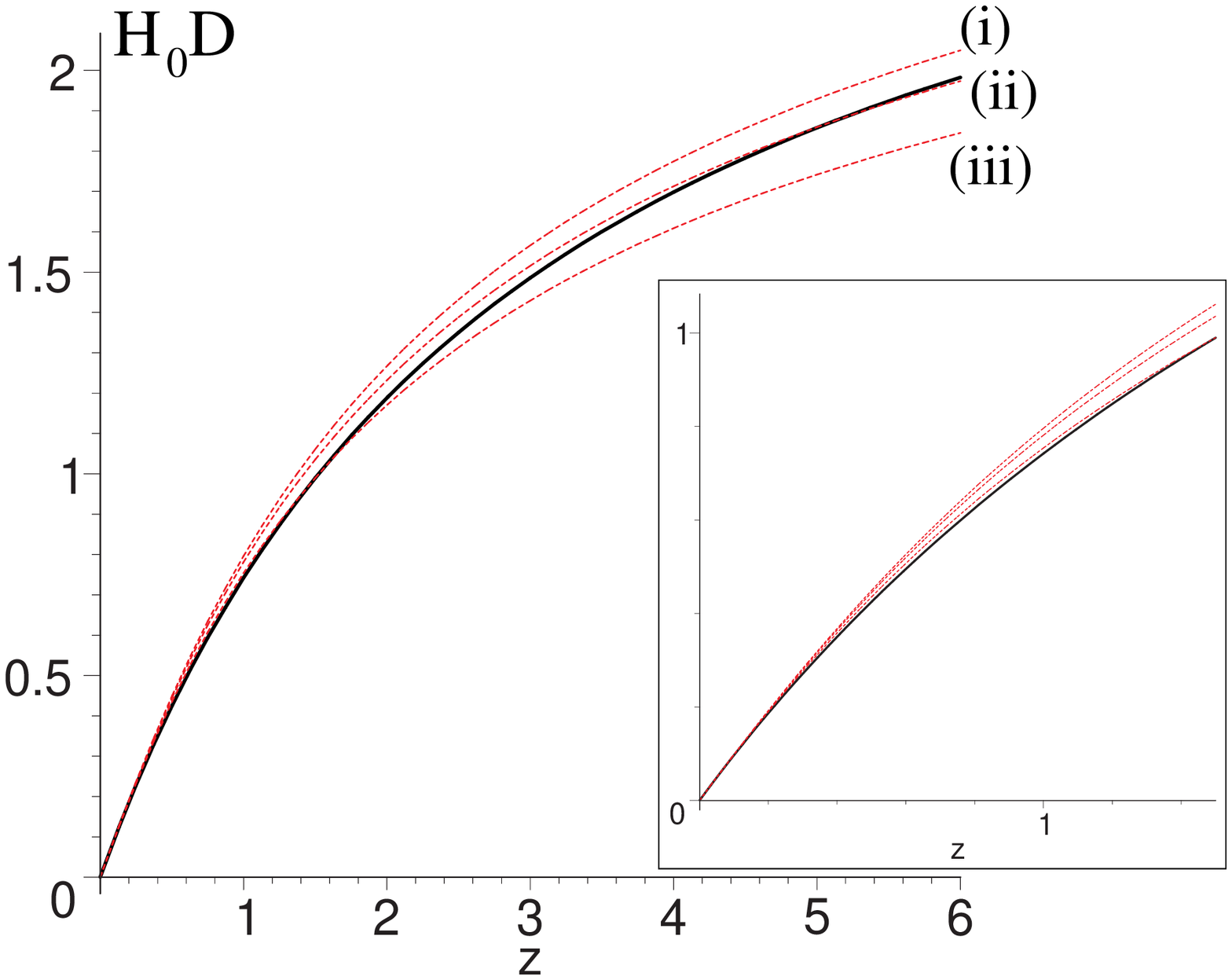}
\qquad\qquad{\sbf(b)}\hskip-15pt
\includegraphics[width=2.in]{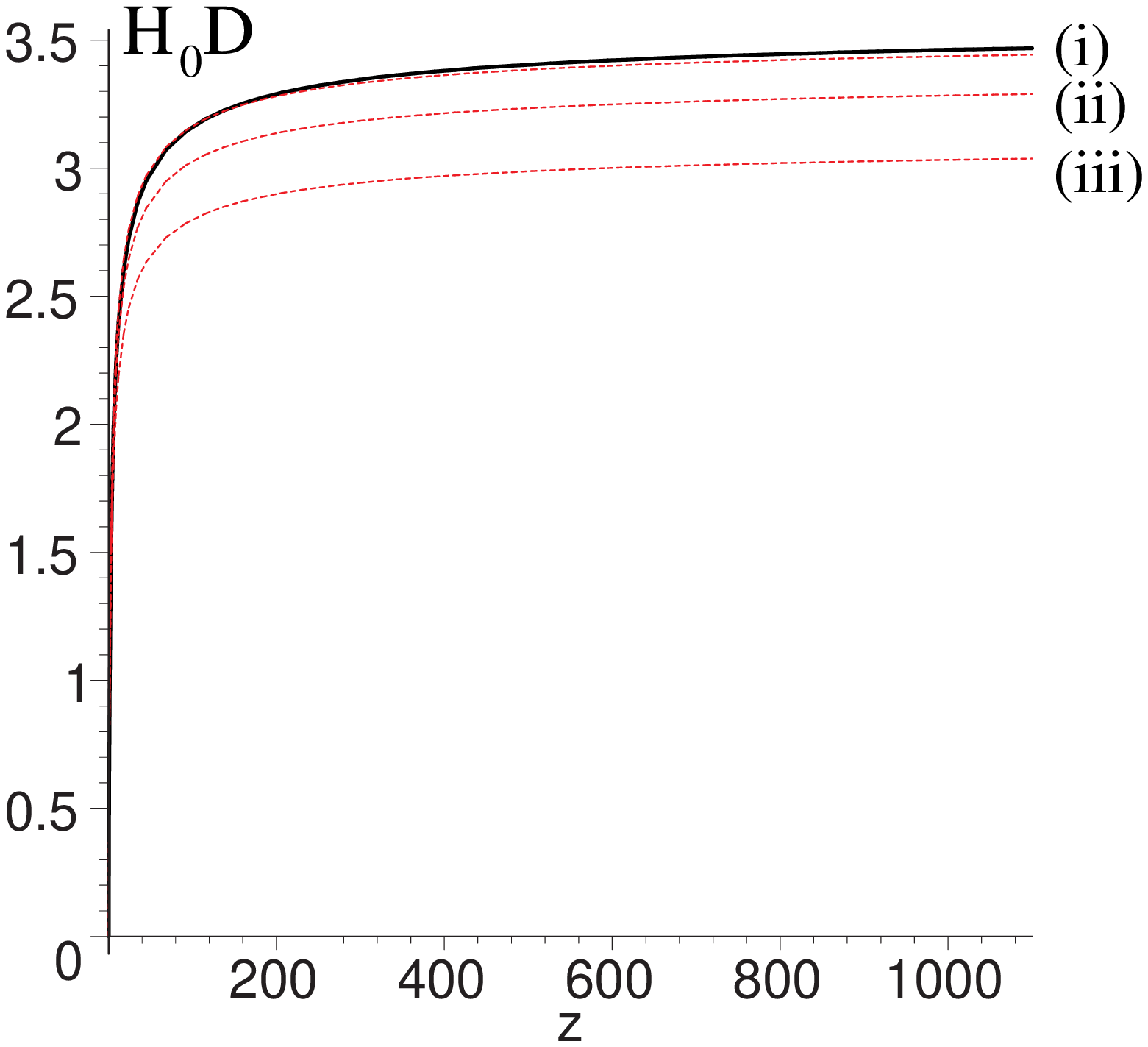}
}
\caption{The effective comoving distance $\Hx D(z)$ is plotted for the
best--fit \name\ (TS) model, with $\fvn=0.762$, (solid line); and for various
spatially flat \LCDM\ models (dashed lines). The parameters for the dashed
lines are (i) $\OmMn=0.249$ (best--fit to WMAP5 only \cite{wmap5}); (ii)
$\OmMn=0.279$ (joint best--fit to SneIa, BAO and WMAP5); (iii)  $\OmMn=0.34$
(best--fit to Riess07 SneIa only \cite{Riess07}). Panel {\bf(a)} shows the
redshift range $z<6$, with an inset for $z<1.5$, which is the range tested by
current SneIa data. Panel {\bf(b)} shows the range $z<1100$ up to the
surface of last scattering, tested by WMAP5.}
\label{fig_coD}
\end{figure}

A direct method of comparing the distance measures with those of homogeneous
models with dark energy, is to observe that for a standard spatially
flat cosmology with dark energy obeying an equation of state $P\Z D=w(z)
\rh\Z D$, the quantity
\beq
\Hm D=\int_0^z{\dd z'\over \sqrt{\OmMn(1+z')^3+\Omega\Z{D0}
\exp\left[3\int_0^{z'}{(1+w(z''))\dd z''\over 1+z''}\right]}}\,,
\label{rFLRW}\eeq
does not depend on the value of the Hubble constant, $\Hm$, but only
directly on $\OmMn=1-\Omega\Z{D0}$. Since the best-fit values of $\Hm$
are potentially different for the different scenarios, a comparison of
$\Hm D$ curves as a function of redshift for the \name\ model versus the
\LCDM\ model gives a good indication of where the largest differences can be
expected, independently of the value of $\Hm$. Such a
comparison is made in Fig.~\ref{fig_coD}.

We see that as redshift increases the \name\ model interpolates between
\LCDM\ models with different values of $\OmMn$. For redshifts $z\lsim1.5$
$\Dfb$ is very close to $\Dlcdm$ for the parameter values $(\OmMn,\OmLn)
=(0.34,0.66)$ (model (iii)) which best--fit the Riess07 supernovae (SneIa) data
\cite{Riess07} only, by our own analysis. For very large
redshifts that approach the surface of last scattering, $z\lsim1100$, on the
other hand, $\Dfb$ very closely matches $\Dlcdm$  for the parameter values
$(\OmMn,\OmLn) =(0.249,0.751)$ (model (i)) which best--fit WMAP5 only
\cite{wmap5}. Over redshifts $2\lsim z\lsim10$, at which scales independent
tests are conceivable, $\Dfb$ makes a transition over corresponding
curves of $\Dlcdm$ with intermediate values of $(\OmMn,\OmLn)$. The
$\Dlcdm$ curve for joint best-fit parameters to SneIa, BAO measurements and
WMAP5 \cite{wmap5}, $(\OmMn,\OmLn) =(0.279,0.721)$ is best--matched over
the range $5\lsim z\lsim 6$, for example.

The difference of $\Dfb$ from any single $\Dlcdm$ curve is perhaps most
pronounced in the range $2\lsim z\lsim 6$, which may be an optimal
regime to probe in future experiments. Gamma--ray bursters
(GRBs) now probe distances to redshifts $z\lsim8.3$, and could be very useful.
A considerable amount work of work has already been done on Hubble diagrams
for GRBs. (See, e.g., \cite{GRB}.) Much more work is needed to nail down
systematic uncertainties, but GRBs may eventually provide a definitive
test in future. An analysis of the \name\ model Hubble diagram using 69 GRBs
has just been performed by Schaefer \cite{Schaefer}, who finds that it fits
the data better than the concordance \LCDM\ model, but not yet by a
huge margin. As more data is accumulated, it should become possible to
distinguish the models if the issues with the standardization of GRBs
can be ironed out.

\subsection{The effective ``equation of state''}

It should be noted that the shape of the $\Hm D$ curves depicted in
Fig.~\ref{fig_coD} represent the observable quantity one is actually
measuring when some researchers loosely talk about ``measuring the
equation of state''. For spatially flat dark energy models, with $\Hm D$
given by (\ref{rFLRW}), one finds that the function $w(z)$ appearing in the
fluid equation of state $P\Z D=w(z)\rh\Z D$ is related to the first
and second derivatives of (\ref{rFLRW}) by
\beq
w(z)={\frn23(1+z)D'^{-1}D''+1\over\OmMn(1+z)^3\Hm^2 D'^2-1}
\label{eos}\eeq
where prime denotes a derivative with respect to $z$. Such a relation
can be applied to observed distance measurements, regardless of whether
the underlying cosmology has dark energy or not. Since it involves
first and second derivatives of the observed quantities, it is actually
much more difficult to determine observationally than directly fitting
$\Hm D(z)$.

\begin{figure}
\centerline{{\sbf(a)}\hskip-5pt
\includegraphics[width=2.in]{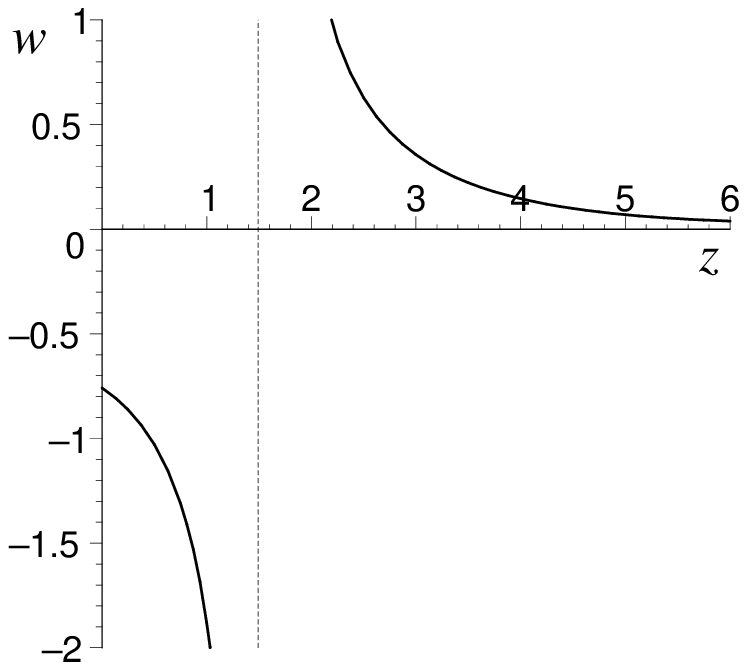}\qquad\qquad
{\sbf(b)}\hskip-5pt
\includegraphics[width=2.in]{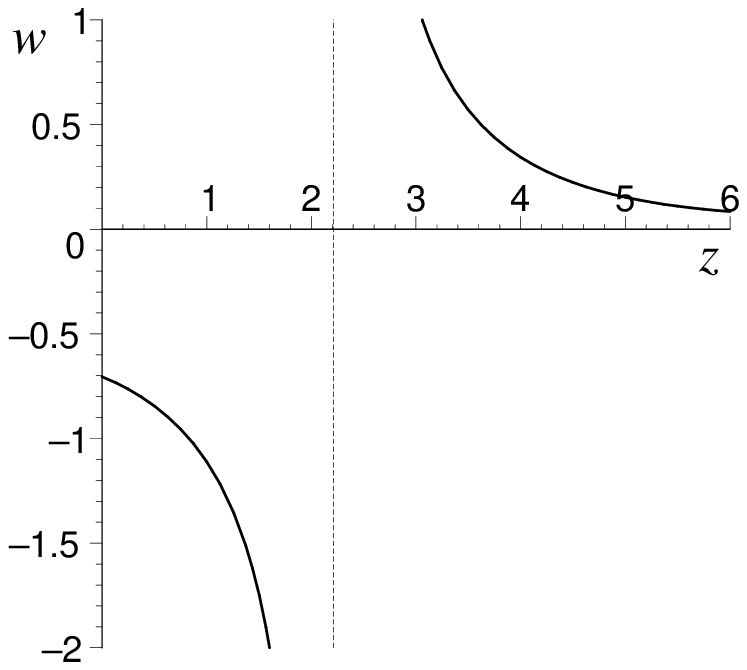}
}
\caption{The artificial equivalent of an equation of state
constructed using the effective comoving distance
(\ref{eos}), plotted for the \name\ tracker solution with best--fit value
$\fvn=0.762$, and two different values of $\OmMn$: {\bf(a)} the canonical
dressed value $\OmMn=\frn12(1-\fvn)(2+\fvn)=0.33$; {\bf(b)} $\OmMn=0.279$.}
\label{fig_wz}
\end{figure}

The equivalent of the ``equation of state'', $w(z)$, for the \name\ model
is plotted in Fig.~\ref{fig_wz}. The fact that $w(z)$ is undefined at
a particular redshift and changes sign through $\pm\infty$ simply reflects
the fact that in (\ref{eos}) we are dividing by a quantity which goes
to zero for the \name\ model, even though the underlying curve of
Fig.~\ref{fig_coD} is smooth. Since one is not dealing with a dark energy
fluid in the present case, $w(z)$ simply has no physical meaning.
Nonetheless, phenomenologically the results
do agree with the usual inferences about $w(z)$ for fits of standard dark
energy cosmologies to SneIa data. For the canonical model of
Fig.~\ref{fig_wz}(a) one finds that the average value of $w(z)\simeq-1$
on the range $z\lsim0.7$, while the average value of $w(z)<-1$ if the
range of redshifts is extended to higher values. The $w=-1$ ``phantom divide''
is crossed at $z\simeq0.46$ for $\fvn\simeq0.76$. One recent study \cite{ZZ}
finds mild 95\% evidence for an equation of state that crosses the phantom
divide from $w>-1$ to $w<-1$ in the range $0.25<z<0.75$ in accord with
the \name\ expectation. By contrast,
another study \cite{SCHMPS} at redshifts $z<1$ draws different conclusions
about dynamical dark energy, but for the given uncertainties in $w(z)$ the
data is consistent with Fig.~\ref{fig_coD}(a) as well as with a cosmological
constant \cite{obs}.

The fact that $w(z)$
is a different sign to the dark energy case for $z>2$ is another
way of viewing our statement above that the redshift range $2\lsim z\lsim6$
may be optimal for discriminating model differences.
\subsection{The $H(z)$ measure}

Further observational diagnostics can be devised if the expansion rate
$H(z)$ can be observationally determined as a function of redshift. Recently
such a determination of $H(z)$ at $z=0.24$ and $z=0.43$ has been made using
redshift space distortions of the BAO scale
in the \LCDM\ model \cite{GCH}. This technique is
of course model dependent, and the Kaiser effect would have to be re-examined
in the \name\ model before a direct comparison of observational results could
be made. A model--independent measure of $H(z)$, the redshift time drift
test, is discussed below.

\begin{figure}
\includegraphics[width=2.in]{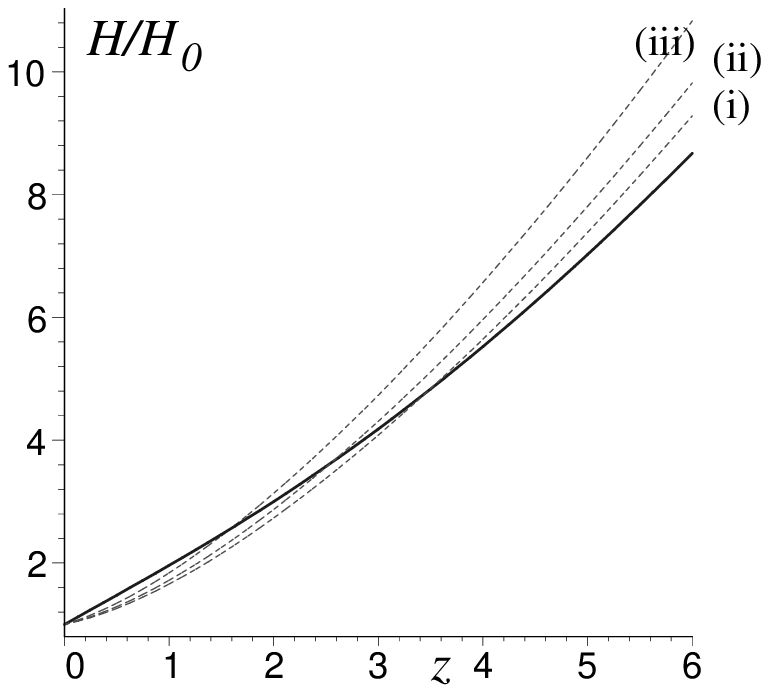}
\caption{The function $\Hm^{-1} H(z)$ for the \name\ model with
$\fvn=0.762$ (solid line) is compared to $\Hm^{-1} H(z)$ for three 
spatially flat \LCDM\ models with the same values of $(\OmMn,\OmLn)$ as in
Fig.~\ref{fig_coD} (dashed lines).}
\label{fig_HH0}
\end{figure}

In Fig.~\ref{fig_HH0} we compare $H(z)/\Hm$ for the \name\ model to
spatially flat \LCDM\ models with the same parameters chosen in
Fig.~\ref{fig_coD}. The most notable feature is that the slope of $H(z)/\Hm$
is less than in the \LCDM\ case, as is to be expected for a model whose
(dressed) deceleration parameter varies more slowly than for \LCDM.
\subsection{The $Om(z)$ measure}

Recently a number of authors \cite{GCC,SSS,ZC} have discussed various
roughly equivalent diagnostics of dark energy. For example, Sahni, Shafieloo
and Starobinsky \cite{SSS}, have proposed a diagnostic function
\beq
Om(z)=\Bigl[{H^2(z)\over\Hm^2}-1\Bigr]\left[(1+z)^3-1\right]^{-1}\,,
\label{dSSS}\eeq
on account of the fact that it is equal to the constant present epoch
matter density parameter, $\OmMn$, at all redshifts for a spatially flat
FLRW model with pressureless dust and a cosmological constant. However, it is
not constant if the cosmological constant is replaced by other forms of dark
energy. For general FLRW models, $H(z)=[D'(z)]^{-1}\sqrt{1+\Omkn\Hm^2 D^2(z)}$,
which only involves a single derivatives of $D(z)$. Thus the diagnostic
(\ref{dSSS}) is easier to reconstruct observationally than the equation
of state parameter, $w(z)$.
\begin{figure}[htb]
\centerline{{\sbf(a)}\hskip-5pt
\includegraphics[width=2.2in]{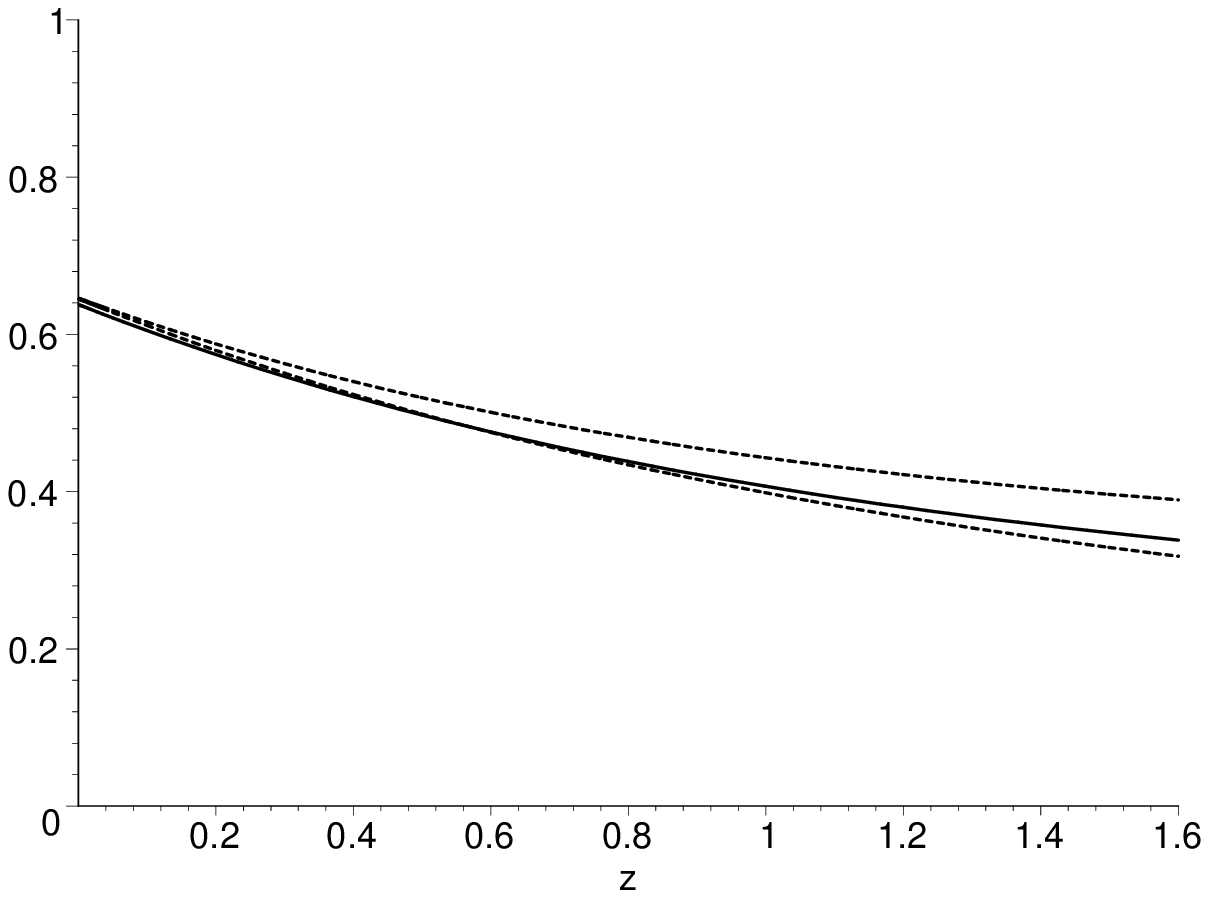}\qquad\qquad
{\sbf(b)}\hskip-5pt
\includegraphics[width=2.2in]{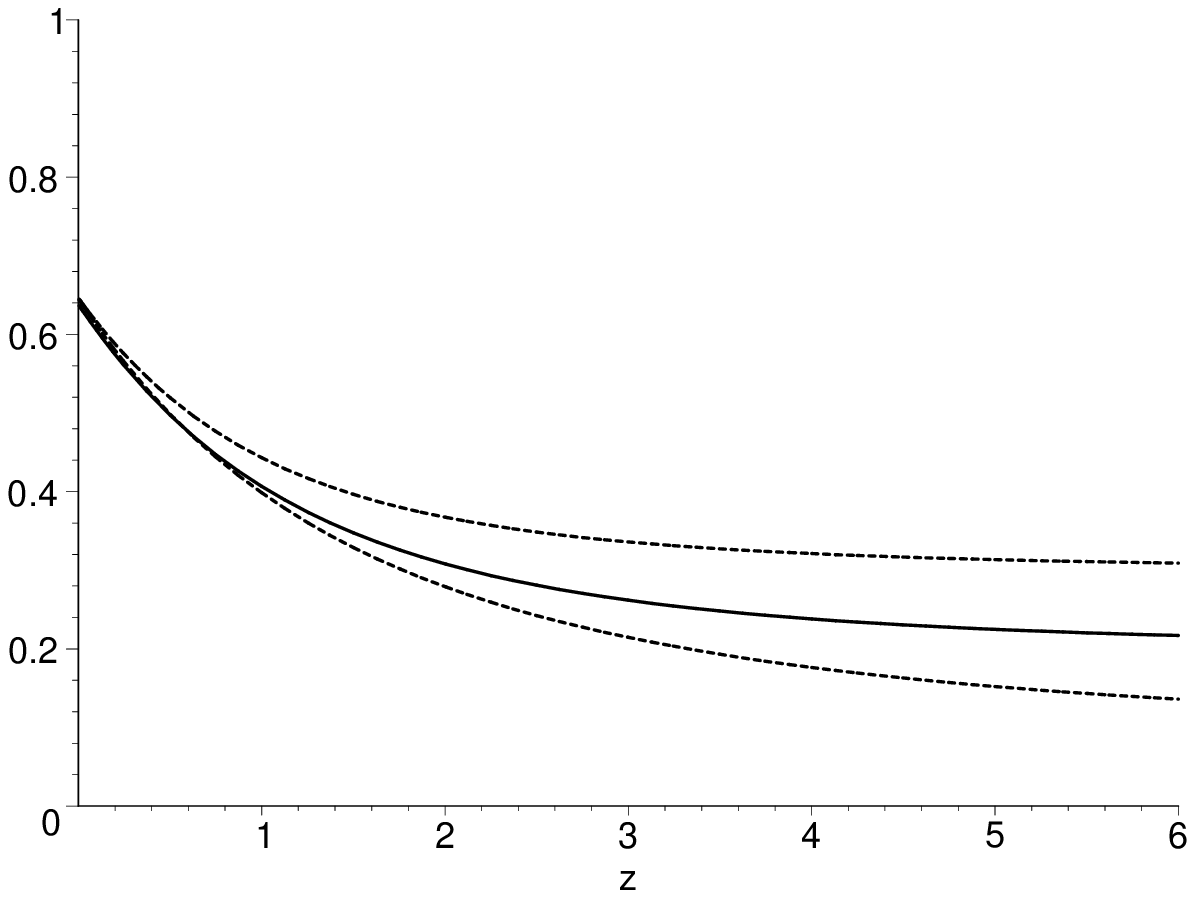}
}
\caption{The dark energy diagnostic $Om(z)$ of Sahni, Shafieloo and Starobinsky
\cite{SSS} plotted for the \name\ tracker solution with best--fit value
$\fvn=0.762$ (solid line), and $1\si$ limits (dashed lines) from ref.\
\cite{LNW}: {\bf(a)} for the redshift range $0<z<1.6$ as shown in
ref.\ \cite{SSS2}; {\bf(b)} for the redshift range $0<z<6$.}
\label{fig_Om}
\end{figure}

The quantity $Om(z)$ is readily calculated for the \name\ model, and the
result is displayed in Fig.~\ref{fig_Om}. What is
striking about Fig.~\ref{fig_Om}, as compared to the curves for quintessence
and phantom dark energy models as plotted in ref.\ \cite{SSS}, is that the
initial value
\beq
Om(0)=\frn23\left.H'\right|_0={2(8\fvn^3-3\fvn^2+4)(2+\fvn)\over(4\fvn^2+\fvn
+4)^2}
\label{intc}\eeq
is substantially larger than in the spatially flat dark energy models.
Furthermore, for the \name\ model $Om(z)$ does not asymptote to the dressed
density parameter $\OmMn$ in any redshift range. For quintessence models
$Om(z)>\OmMn$,
while for phantom models $Om(z)<\OmMn$, and in both cases $Om(z)\to\OmMn$
as $z\to\infty$. In the \name\ model, $Om(z)>\OmMn\simeq0.33$ for $z\lsim1.7$,
while $Om(z)<\OmMn$ for $z\gsim1.7$. It thus behaves more like a quintessence
model for low $z$, in accordance with Fig.~\ref{fig_wz}. However, the
steeper slope and the different large $z$ behaviour mean the
diagnostic is generally very different to that of typical dark energy models.
For large $z$, $\OMMn<Om(\infty)<\OmMn$, if $\fvn>0.25$.

Interestingly enough, a recent analysis of SneIa, BAO and CMB data
\cite{SSS2} for dark energy models
with two different empirical fitting functions for $w(z)$ gives an intercept
$Om(0)$ which is larger than expected for typical quintessence or phantom
energy models, and in the better fit of the two models the intercept (see
Fig.~3 of ref.\ \cite{SSS2}) is close to the value expected for the \name\
model, which is tightly constrained to the range $0.638<Om(0)<0.646$ if
$\fvn=0.76^{+0.12}_{-0.09}$.

\subsection{The Alcock--Paczy\'nski test and baryon acoustic oscillations}

Some time ago Alcock and Paczy\'nski devised a test \cite{AP} which relies on
comparing the radial and transverse proper length scales of spherical standard
volumes comoving with the Hubble flow. This test, which determines the function
\beq
f\Ns{AP}={1\over z}\left|\Deriv\de z\th\right|={HD\over z},
\label{fAP}
\eeq
was originally conceived to distinguish FLRW models with a cosmological
constant from those without a $\Lambda$ term. The test is free from many
evolutionary effects, but relies on one being able to remove systematic
distortions due to peculiar velocities.

Current detections of the BAO scale in galaxy clustering statistics
\cite{bao,Percival} can in fact be viewed as a variant of the
Alcock--Paczy\'nski test, as they make use of both the transverse and
radial dilations of the fiducial comoving BAO scale to present a measure
\beq
D\Z V=\left[zD^2\over H(z)\right]^{1/3}=Df\Ns{AP}^{-1/3}.
\label{BAOr}\eeq
\begin{figure}
\centerline{{\sbf(a)}\hskip-5pt
\includegraphics[width=1.8in]{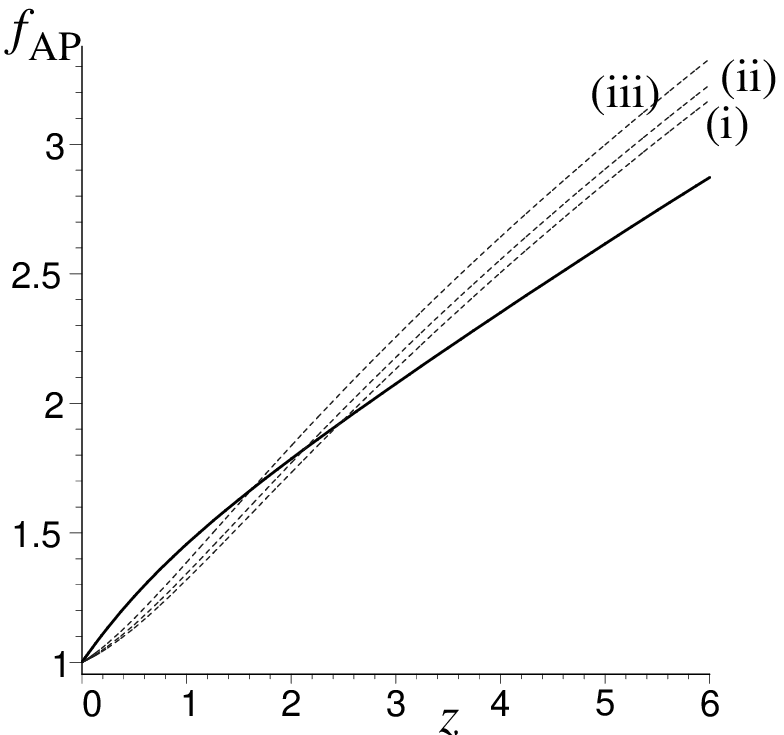}\qquad\qquad
{\sbf(b)}\hskip-5pt
\includegraphics[width=2.in]{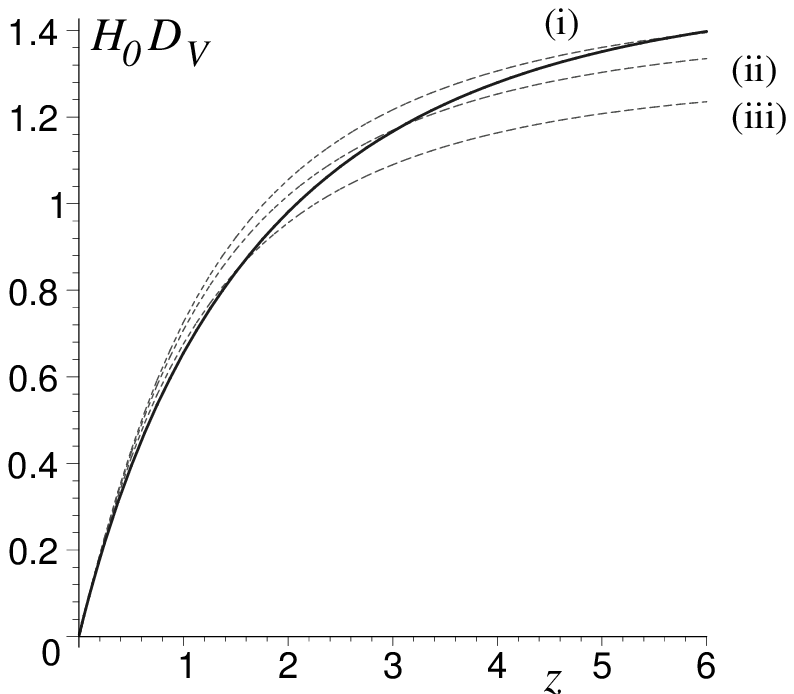}
}
\caption{{\bf(a)} The Alcock--Paczy\'nski test function $f\Ns{AP}=HD/z$;
and {\bf(b)} the BAO radial test function $\Hm D\Z V=\Hm Df\Ns{AP}^{-1/3}$.
In each case the \name\ model with
$\fvn=0.762$ (solid line) is compared to three 
spatially flat \LCDM\ models with the same values of $(\OmMn,\OmLn)$ as in
Fig.~\ref{fig_coD} (dashed lines).}
\label{fig_AP}
\end{figure}

In Fig.~\ref{fig_AP} the Alcock--Paczy\'nski test function (\ref{fAP})
and BAO scale measure (\ref{BAOr}) of the \name\ model are
compared to those of the spatially flat \LCDM\ model with different values of
($\OmLn$,$\OmLn$). Over the range of redshifts $z<1$ studied currently
with galaxy clustering statistics, the $f\Ns{AP}$ curve distinguishes
the \name\ model from the \LCDM\ models much more strongly than the
$D\Z V$ test function. In particular, the \name\ $f\Ns{AP}$ has a distinctly
different shape to that of the \LCDM\ model, being convex. The primary
reason for use of the integral measure (\ref{BAOr}) has been a lack of data.
Future measurements with enough data to separate the radial and angular BAO
scales are a potentially powerful way of distinguishing the \name\ model from
\LCDM.

Recently Gazta\~naga, Cabr\'e and Hui \cite{GCH} have made the first
efforts to separate the radial and angular BAO scales in different redshift
slices. Although they have not yet published separate values for the radial
and angular scales, their results are interesting when compared to the
expectations of the \name\ model. Their study yields best-fit values
of the present total matter and baryonic matter density parameters,
$\OmMn$ and $\OmBn$, which are in tension with WMAP5 parameters fit to
the \LCDM\ model. In particular, the ratio of non-baryonic cold dark
matter to baryonic matter has a best-fit value $\OmCn/\OmBn=(\OmMn-\OmBn)/
\OmBn$ of 3.7 in the $0.15<z<0.3$ sample, 2.6 in the $0.4<z<0.47$ sample,
and 3.6 in the whole sample, as compared to the expected value of 6.1 from
WMAP5. The analysis of the 3--point correlation function yields similar 
conclusions, with a best fit \cite{GCCCF} $\OmMn=0.28\pm0.05$,
$\OmBn=0.079\pm0.025$. By comparison, the parameter fit to the \name\ model
of ref.\ \cite{LNW} yields dressed parameters $\OmMn=0.33^{+0.11}_{-0.16}$,
$\OmBn=0.080^{+0.021}_{-0.013}$, and a ratio $\OmCn/\OmBn=3.1^{+2.5}_{-2.4}$. 
Since other forms of dark energy are not generally expected to give rise
to a renormalization of the ratio of non-baryonic to baryonic matter, this
is encouraging for the \name\ model.

\subsection{Test of (in)homogeneity}

Recently Clarkson, Bassett and Lu \cite{CBL} have constructed what they call
a ``test of the Copernican principle'' based on the observation that
for homogeneous, isotropic models which obey the Friedmann equation,
the present epoch curvature parameter, a constant, may be written as
\beq
\Omkn={[H(z)D'(z)]^2-1\over[\Hm D(z)]^2}\label{ctest1}
\eeq
for all $z$, irrespective of the dark energy model or any other model
parameters. Consequently, taking a further derivative, the quantity
\beq
\CC(z)\equiv1+H^2(DD''-D'^2)+HH'DD'\label{ctest2}
\eeq
must be zero for all redshifts for any FLRW geometry.

A deviation of $\CC(z)$ from zero, or of (\ref{ctest1}) from a constant
value, would therefore mean that the assumption of homogeneity is violated.
Although this only constitutes a test of the assumption of the Friedmann
equation, i.e., of the Cosmological Principle rather than the broader
Copernican Principle adopted in ref.\ \cite{clocks}, the average
inhomogeneity will give a clear and distinct prediction of a non-zero
$\CC(z)$ for the \name\ model.
\begin{figure}
{\sbf(a)}
\includegraphics[width=2.2in]{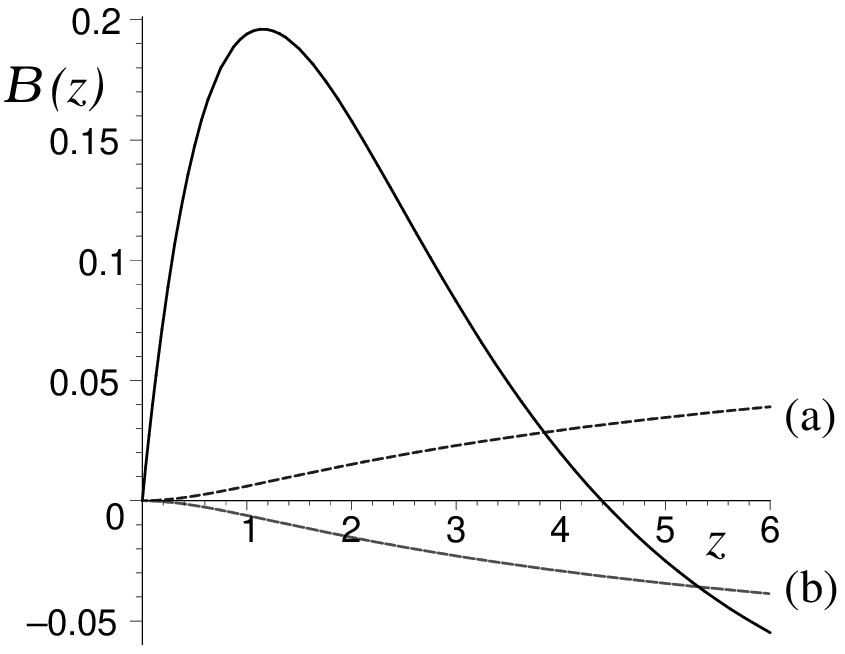}\qquad\qquad
{\sbf(b)}
\includegraphics[width=2.in]{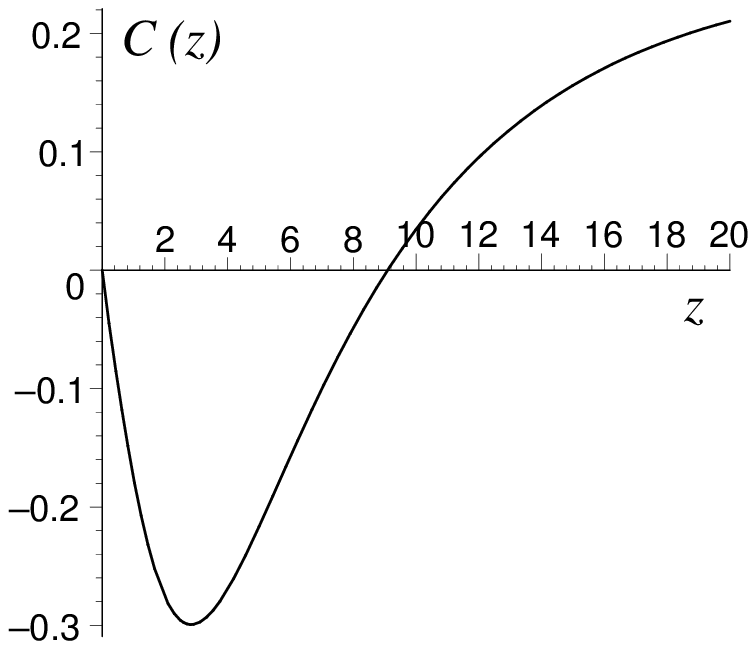}
\caption{{\bf Left panel:}
The (in)homogeneity test function $\BB(z)=[HD']^2-1$ is plotted for
the \name\ tracker solution with best--fit value $\fvn=0.762$ (solid line), and
compared to the equivalent curves $\BB=\Omkn(\Hm D)^2$ for two different
\LCDM\ models with small curvature:\break {\bf(a)} $\OmMn=0.28$, $\OmLn=0.71$,
$\Omkn=0.01$; {\bf(b)} $\OmMn=0.28$, $\OmLn=0.73$, $\Omkn=-0.01$.\hfil\break
{\bf Right panel:} The (in)homogeneity test function $\CC(z)$ is plotted for
the $\fvn=0.762$ tracker solution.}
\label{fig_Bex}
\end{figure}

The functions (\ref{ctest1}) and (\ref{ctest2}) are computed in ref.\
\cite{obs}.
Observationally it is more feasible to fit (\ref{ctest1}) which involves one
derivative less of redshift. In Fig.\ \ref{fig_Bex} we exhibit both
$\CC(z)$, and also the function
$\BB(z)=[HD']^2-1$ from the numerator of (\ref{ctest1}) for the \name\
model, as compared to two \LCDM\ models with a small amount of spatial
curvature. A spatially flat FLRW model would have $\BB(z)\equiv0$. In other
FLRW cases $\BB(z)$ is always a monotonic
function whose sign is determined by that of $\Omkn$. An open $\Lambda=0$
universe with the same $\OmMn$ would have a monotonic function $\BB(z)$
very much greater than that of the \name\ model.

\subsection{Time drift of cosmological redshifts}

For the purpose of the $Om(z)$ and (in)homogeneity tests considered in the
last section, $H(z)$ must be observationally determined, and this is
difficult to achieve in a model independent way. There is one way of
achieving this, however, namely by measuring the time variation of the
redshifts of different sources over a sufficiently long time interval
\cite{SML}, as has been discussed recently by Uzan, Clarkson and Ellis
\cite{UCE}. Although the measurement is extremely challenging, it may be
feasible over a 20 year period by precision measurements of the Lyman-$\al$
forest in the redshift range $2<z<5$ with the next generation of
Extremely Large Telescopes \cite{ELT}.

\begin{figure}[htb]
\includegraphics[width=2.2in]{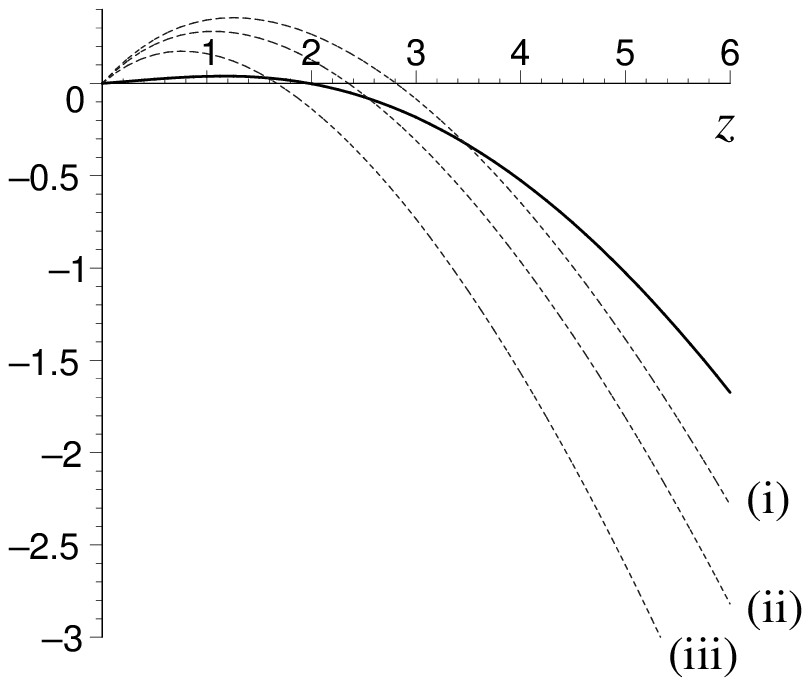}
\caption{The function $\Hm^{-1}\Deriv\dd\ta z$ for the \name\ model with
$\fvn=0.762$ (solid line) is compared to $\Hm^{-1}\Deriv\dd\ta z$ for three 
spatially flat \LCDM\ models with the same values of $(\OmMn,\OmLn)$ as in
Fig.~\ref{fig_coD} (dashed lines).}
\label{fig_zdot}
\end{figure}

In ref.\ \cite{obs} an analytic expression for $\Hm^{-1}\Deriv\dd\ta z$
is determined, the derivative being with respect to wall time for observers
in galaxies. The resulting function is displayed in Fig.~\ref{fig_zdot} for
the best-fit \name\ model with $\fvn=0.762$, where it is compared to the
equivalent function for three different spatially flat \LCDM\ models. What
is notable is that the curve for the \name\ model is considerably flatter
than those of the \LCDM\ models. This may be understood to arise from
the fact that the magnitude of the apparent acceleration is considerably
smaller in the \name\ model, as compared to the magnitude of the acceleration
in \LCDM\ models. For models in which there is no apparent acceleration
whatsoever, one finds that $\Hm^{-1}\Deriv\dd\ta z$ is always negative.
If there is cosmic acceleration, real or apparent, at late epochs then
$\Hm^{-1}\Deriv\dd\ta z$ will become positive at low redshifts, though
at a somewhat larger redshift than that at which acceleration is deemed
to have begun.

Fig.~\ref{fig_zdot} demonstrates that a very clear signal of differences in
the redshift time drift between the \name\ model and \LCDM\ models might
be determined at low redshifts when $\Hm^{-1}\Deriv\dd\ta z$ should be
positive. In particular, the magnitude of $\Hm^{-1}\Deriv\dd\ta z$ is
considerably smaller for the \name\ model as compared to \LCDM\ models.
Observationally, however, it is expected that measurements
will be best determined for sources in the Lyman $\al$ forest in the range,
$2<z<5$. At such redshifts the magnitude of the drift is somewhat
more pronounced in the case of the \LCDM\ models. For a source at $z=4$,
over a period of $\de\ta=10$ years we would have $\de z=-3.3\times10^{-10}$
for the \name\ model with $\fvn=0.762$ and $\Hm=61.7\kmsMpc$. By comparison,
for a spatially flat \LCDM\ model with $\Hm=70.5\kmsMpc$ a source
at $z=4$ would over ten years give $\de z=-4.7\times10^{-10}$ for
$(\OmMn,\OmLn)=(0.249,0.751)$, and $\de z=-7.0\times10^{-10}$ for
$(\OmMn,\OmLn)=(0.279,0.721)$.

\section{Discussion}

The tests outlined here demonstrate several lines of investigation to
distinguish the \name\ model from models of homogeneous dark energy.
The (in)homogeneity test of Bassett, Clarkson and Lu is definitive,
since it tests the validity of the Friedmann equation directly.

In performing these tests, however, one must be very careful to ensure
that data has not been reduced with built--in assumptions that use the
Friedmann equation. For example, current estimates of the BAO scale such as
that of Percival \etal\ \cite{Percival} do not determine $D\Z V$
directly from redshift and angular diameter measures, but first
perform a Fourier space transformation to a power spectrum, assuming
a FLRW cosmology. Redoing such analyses for the \name\ model may
involve a recalibration of relevant transfer functions.

In the case of supernovae, one must also take care as compilations such
as the Union \cite{union} and Constitution \cite{Hicken} datasets
use the SALT method to calibrate light curves. In this approach empirical
light curve parameters and cosmological parameters -- {\it assuming the
Friedmann equation} -- are simultaneously fit by analytic marginalisation
before the raw apparent magnitudes are recalibrated. As Hicken \etal\ discuss
\cite{Hicken}, a number of systematic discrepancies exist between
data reduced by the SALT, SALT2, MLCS31 and MLCS17 techniques even within
the \LCDM\ model. In the case of the \name\ model, we find considerable
differences between the different approaches \cite{SW}. In principle,
at present there appear to be enough supernovae to decide between the \LCDM\
and \name\ models on Bayesian evidence, but one is led to different
conclusions depending on how the data is reduced. It is therefore important
that the systematic issues are unravelled.

The value of the dressed Hubble constant is also an observable quantity of
considerable interest. A recent determination of $\Hm$ by Riess \etal\
\cite{shoes} poses a challenge for the \name\ model. However, it is a
feature of the \name\ model that a 17--22\% variance in the apparent Hubble
flow will exist on local scales below the scale of statistical homogeneity,
and this may potentially complicate calibration of the cosmic distance
ladder. Further quantification of the variance in the apparent Hubble
flow in relationship to local cosmic structures would provide an
interesting possibility for tests of the \name\ cosmology for which there
are no counterparts in the standard cosmology.

\begin{theacknowledgments} I thank Prof.\ Remo Ruffini and
ICRANet for support and hospitality while the work of ref.\ \cite{obs}
was undertaken. This work was also partly supported by the Marsden fund
of the Royal Society of New Zealand.
\end{theacknowledgments}

\end{document}